\documentclass[twocolumn,showpacs,prc,superscriptaddress,amsmath,amssymb]{revtex4-1}

\usepackage{multirow}
\usepackage{graphicx}
\usepackage{dcolumn}
\usepackage{bm}
\usepackage{soul}
\usepackage{color}
\begin{document}

\preprint{APS/123-QED}

\title{Population of $^{13}$Be in a Nucleon Exchange Reaction}

\author{B. R. Marks}
\affiliation{Department of Physics, Hope College, Holland, Michigan 49422, USA}
\author{P. A. DeYoung}
\email{deyoung@hope.edu}
\affiliation{Department of Physics, Hope College, Holland, Michigan 49422, USA}
\author{J. K. Smith}
\altaffiliation[Present address: ]{TRIUMF, 4004 Wesbrook Mall, Vancouver, British Columbia, V6T 2A3 Canada}
\affiliation{National Superconducting Cyclotron Laboratory, Michigan State University, East Lansing, Michigan 48824, USA}
\affiliation{Department of Physics $\&$ Astronomy, Michigan State University, East Lansing, Michigan 48824, USA}
\author{T. Baumann}
\affiliation{National Superconducting Cyclotron Laboratory, Michigan State University, East Lansing, Michigan 48824, USA}
\author{J. Brown}
\affiliation{Department of Physics, Wabash College, Crawfordsville, IN 47933,  USA}
\author{N. Frank}
\affiliation{Department of Physics and Astronomy, Augustana College, Rock Island, IL 61201, USA}
\author{J. Hinnefeld}
\affiliation{Department of Physics and Astronomy, Indiana University at South Bend, South Bend, IN 46634, USA}
\author{M. Hoffman}
\affiliation{Department of Physics and Astronomy, Augustana College, Rock Island, IL 61201, USA}
\author{M. D. Jones}
\affiliation{National Superconducting Cyclotron Laboratory, Michigan State University, East Lansing, Michigan 48824, USA}
\affiliation{Department of Physics $\&$ Astronomy, Michigan State University, East Lansing, Michigan 48824, USA}
\author{Z. Kohley}
\affiliation{National Superconducting Cyclotron Laboratory, Michigan State University, East Lansing, Michigan 48824, USA}
\affiliation{Department of Chemistry, Michigan State University, East Lansing, Michigan 48824, USA}
\author{A. N. Kuchera}
\affiliation{National Superconducting Cyclotron Laboratory, Michigan State University, East Lansing, Michigan 48824, USA}
\author{B. Luther}
\affiliation{Department of Physics, Concordia College, Moorhead, Minnesota 56562, USA}
\author{A. Spyrou}
\affiliation{National Superconducting Cyclotron Laboratory, Michigan State University, East Lansing, Michigan 48824, USA}
\affiliation{Department of Physics $\&$ Astronomy, Michigan State University, East Lansing, Michigan 48824, USA}
\author{S. Stephenson}
\affiliation{Department of Physics, Gettysburg College, Gettysburg, Pennsylvania 17325, USA}
\author{C. Sullivan}
\affiliation{National Superconducting Cyclotron Laboratory, Michigan State University, East Lansing, Michigan 48824, USA}
\affiliation{Department of Physics $\&$ Astronomy, Michigan State University, East Lansing, Michigan 48824, USA}
\author{M. Thoennessen}
\affiliation{National Superconducting Cyclotron Laboratory, Michigan State University, East Lansing, Michigan 48824, USA}
\affiliation{Department of Physics $\&$ Astronomy, Michigan State University, East Lansing, Michigan 48824, USA}
\author{N. Viscariello}
\affiliation{Department of Physics and Astronomy, Augustana College, Rock Island, IL 61201, USA}
\author{S. J. Williams}
\affiliation{National Superconducting Cyclotron Laboratory, Michigan State University, East Lansing, Michigan 48824, USA}

\begin{abstract}
The neutron-unbound nucleus $^{13}$Be was populated with a nucleon-exchange reaction from a 71~MeV/u secondary $^{13}$B beam. The decay energy spectrum was reconstructed using invariant mass spectroscopy based on $^{12}$Be fragments in coincidence with neutrons. The data could be described with an $s$-wave resonance at $E_r$ = 0.73(9)~MeV with a width of $\Gamma_r$ = 1.98(34)~MeV and a $d$-wave resonance at $E_r$ = 2.56(13)~MeV with a width of $\Gamma_r$ = 2.29(73)~MeV. The observed spectral shape is consistent with previous one-proton removal reaction measurements from $^{14}$B.
\end{abstract}

\pacs{21.10.Pc, 25.60.-t, 27.20+n, 29.30.Hs}

\maketitle

\section{Introduction}

Recent experimental investigations of the level structure of the neutron-unbound nucleus $^{13}$Be agree about the overall strength distribution of the excitation energy spectrum \cite{Lec04,Sim07,Kon10,Aks13a,Aks13b,Ran14}, but there is no consensus on its interpretation. While there seems to be general agreement about the presence of a broad $s$-wave resonance below 1 MeV and a $d$-wave resonance at 2 MeV, the composition of the observed peak around 500 keV, as well as the decay paths of the $d$-wave resonance, are still being discussed. Earlier reports of a narrow low-lying $s$-wave state \cite{Tho00,Chr08} have been attributed to a sequential decay from the first excited 2$^+$ state in $^{14}$Be to $^{12}$Be \cite{Kon10,Ran14,Bau12}. 

In 2010, Kondo {\it et al.}~\cite{Kon10} reported a low-lying $p$-wave resonance at 510(10)~keV populated by a one-neutron removal reaction from $^{14}$Be at 69 MeV/u. However, a recent analysis of these data, as well as a new measurement at a higher beam energy on a hydrogen target (304 MeV/u), preferred an interpretation which fits the $\sim$500 keV peak with only two interfering broad $s$-wave resonances \cite{Aks13a,Aks13b}. Moreover, the presence of additional $p$- or $d$-wave strength could not be ruled out, indicating that an $\ell \neq 0$ resonance around 1 MeV might exist \cite{Aks13b}. The fits in both papers included a significant decay branch of the $d_{5/2}$ state to the first excited 2$^+$ state in $^{12}$Be.

While neutron-removal reactions are expected to populate positive as well as negative-parity states, proton-removal reactions should be more selective and populate only positive-parity states. Randisi {\it et al.} \cite{Ran14} measured the decay energy spectrum of $^{13}$Be following the one-proton removal reaction from $^{14}$B at 35 MeV/u and argued that the $\sim$500 keV peak consists of an $s$-wave resonance as well as a low-lying $d$-wave resonance. In addition, Randisi {\it et al.} searched for the decay of the $d_{5/2}$ resonance at 2 MeV to the first excited 2$^+$ state in $^{12}$Be by measuring the $\gamma$-rays from this state in coincidence. No significant branch of this decay mode was observed.

In the present work, the nucleon-exchange reaction ($-$1p+1n) from $^{13}$B was used to populate states in $^{13}$Be. Similar to the proton-removal reaction it is expected to only populate positive-parity states. This type of reaction has been shown to have sizable cross sections at intermediate beam energies. For example,  the one-proton removal--one-neutron addition ($-$1p+1n) reaction has been utilized with stable ($^{48}$Ca) as well as radioactive ($^{48}$K and $^{46}$Cl) beams to explore the structures of $^{48}$K, $^{48}$Ar, and $^{46}$S \cite{Gad09}. The inclusive cross sections were 0.13(1)~mb and 0.057(6)~mb for the $^9$Be($^{48}$K,$^{48}$Ar) and $^9$Be($^{46}$Cl,$^{46}$S), respectively. This ($-$1p+1n) reaction was also used for the first time to measure neutron unbound states in the study of $^{26}$F populated from a 86 MeV/u $^{26}$Ne beam \cite{Fra11}.

\section{Experimental Setup}

The experiment was performed at the National Superconducting Cyclotron Laboratory at Michigan State University. A 120 MeV/u $^{18}$O primary beam from the Coupled Cyclotron Facility bombarded a 2.5~g/cm$^2$ $^9$Be production target.  The A1900 fragment separator was used to separate and select the $^{13}$B secondary beam. The final energy of the beam was 71 MeV/u, with an intensity of approximately 8$\times$10$^5$ particles per second and a purity of 96\%. The $^{13}$B beam impinged upon a 51~mg/cm$^2$ $^9$Be target where $^{13}$Be was produced in a nucleon-exchange reaction and immediately decayed into $^{12}$Be + n.

The $^{12}$Be reaction products were deflected by a large-gap sweeper magnet \cite{Bir05} and identified from energy-loss and time-of-flight measurements. The $^{12}$Be energy and momentum vectors were reconstructed from position information and a transformation matrix based on the magnetic-field map using the program {\sc Cosy} Infinity~\cite{Mak05}. Coincident neutrons were measured with the Modular Neutron Array (MoNA)~\cite{Lut03,Bau05} and the Large-area multi-Institutional Scintillator Array (LISA). The energy and momentum vectors of the neutrons were determined from the positions of the neutron interactions in the arrays and the time-of-flight between the arrays and a scintillator located upstream near the target. The nucleon-exchange data were recorded simultaneously with the data for the one proton-removal reaction populating unbound states in $^{12}$Be. These results have been published recently in Ref. \cite{Smi14} where further details of the experimental setup and analysis can be found.

\section{Data Analysis}

\begin{figure}[tb]
	\centering
	\includegraphics[width=0.48\textwidth]{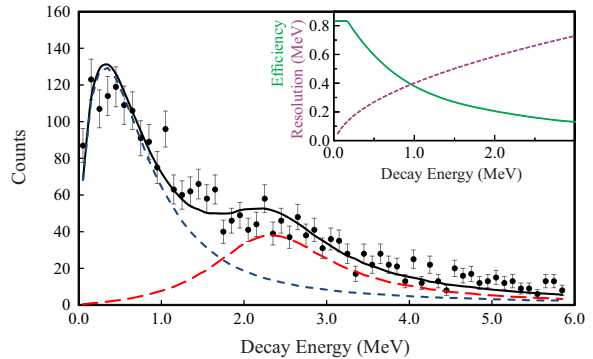}
	\caption{(color online) Decay-energy spectrum of $^{13}$Be fit with two components.  The solid black line is the sum of simulated decay-energy spectra from an {\it s}-wave resonance (short-dashed blue line) and a {\it d}-wave resonance (long-dashed red line) with parameters listed in the text. The insert shows the energy dependent resolution (dotted purple line) and the overall efficiency (solid green line).}
\label{Figure1}
\end{figure}

\begin{figure}[tb]
	\centering
	\includegraphics[width=0.48\textwidth]{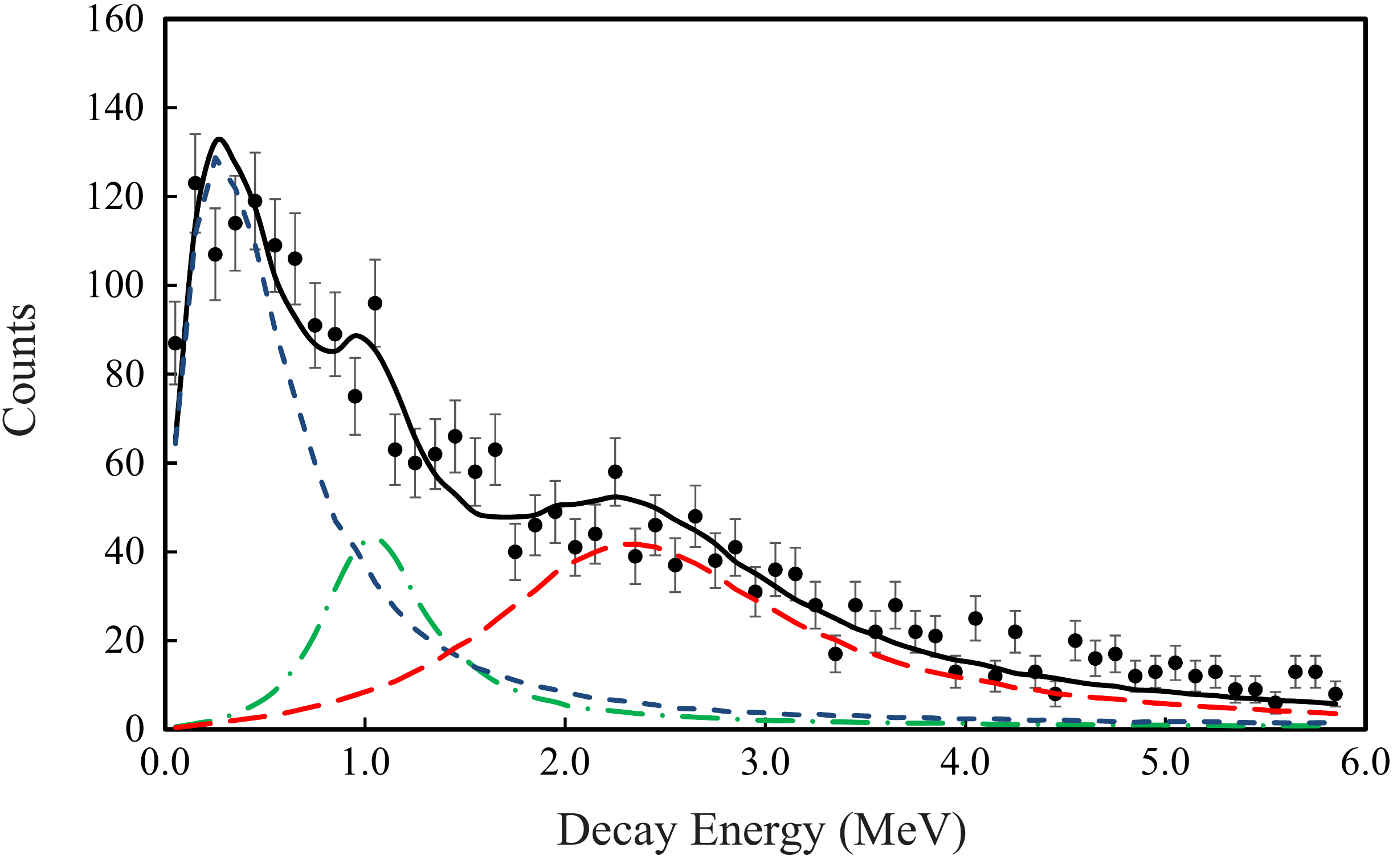}
	\caption{(color online)  Decay-energy spectrum of $^{13}$Be fit with three components. The solid black line is the sum of simulated decay-energy spectra from an {\it s}-wave resonance (short-dashed red line) and two {\it d}-wave resonances (long-dashed red line and dot-dashed green line) with parameters listed in the text.}
\label{Figure2}
\end{figure}

The decay energy spectrum of $^{13}$Be was reconstructed by the invariant-mass method and is shown in Figures \ref{Figure1} and \ref{Figure2}. The spectrum shows the same general features as the previous measurements with a strong peak around 500 keV and an additional structure at about 2 MeV. The energy dependent resolution (blue-dotted line) and the overall efficiency (red solid line) are shown in the insert of Figure \ref{Figure1}.

In order to interpret the measured decay-energy spectrum, Monte Carlo simulations were performed with the incoming beam characteristics, reaction mechanism, and detector resolutions taken into account. The neutron interactions within MoNA-LISA were simulated with G{\sc eant}4~\cite{Ago03,All06} using the {\sc menate\_r} package \cite{Roe08} as described in Ref.~\cite{Koh12}.  Resonances were parameterized using energy-dependent Breit-Wigner line shapes \cite{Smi14}.

The present nucleon-exchange reaction is expected to populate the same positive-parity states that were populated in the one-proton removal reaction. In that case, the valence neutron configuration of the $^{14}$B projectile is dominated by $\nu 2s_{1/2}$ and $\nu 1d_{5/2}$ components and states with the same configurations are expected to be populated in $^{13}$Be by proton removal \cite{Ran14}. The ground state of $^{13}$B has spin and parity of 3/2$^-$ dominated by a ($\pi 1p_{3/2}$)$^3$ proton configuration and a closed $sp$ shell neutron configuration. Removing the odd proton from $^{13}$B is similar to the proton removal from $^{14}$B while the added extra odd neutron will populate states in the open $sd$ shell.

Randisi {\it et al.} was able to fit their data from the proton-removal reaction based on selectivity arguments with only two components, an $s$-wave resonance at $E_r$ = 0.70(11)~MeV with a width of $\Gamma_r$ = 1.70(22)~MeV and a $d$-wave resonance at $E_r$ = 2.40(14)~MeV with a width of $\Gamma_r$ = 0.70(32)~MeV \cite{Ran14}. The best fit to the decay-energy spectrum from the present nucleon-exchange reactions is shown in Figure \ref{Figure1} with an $s$-wave resonance at $E_r$ = 0.73(9)~MeV with a width of $\Gamma_r$ = 1.98(34)~MeV and a $d$-wave resonance at $E_r$ = 2.56(13)~MeV with a width of $\Gamma_r$ = 2.29(73)~MeV. Overall these parameters agree with the results from Randisi {\it et al.} with only the width of the $d$-wave resonance being somewhat larger.

The overall cross section for populating $^{13}$Be with the ($-$1p+1n) reaction was extracted to be 0.30(15)~mb which is about an order of magnitude smaller than one-proton removal reactions on neutron-rich $p$-shell nuclei. Kryger {\it et al.} reported a cross section of 2.46(3)~mb for the proton removal from $^{16}$C to $^{15}$B \cite{Kry96} and Lecouey {\it et al.} measured 6.5(15)~mb for the proton removal recation from $^{17}$C to $^{16}$B \cite{Lec09}. 

The cross section is somewhat larger than the cross section of 0.1~mb estimated for the charge-exchange reaction based on Distorted Wave Born Approximation (DWBA) calculations using the code F{\sc OLD} \cite{Coo88}.  Transition densities that were input to F{\sc OLD}  were calculated using the shell-model code O{\sc XBASH} \cite{Bro04}. The CKII interaction \cite{Coh65} was used in the $p$-shell model space to calculate the transition densities for the $^9$Be--$^9$B system, and the WBP interaction \cite{War92} was used in the {\it spsdpf}-shell model space to calculate the transition densities for the $^{13}$B--$^{13}$Be system. The effective nucleon--nucleon interaction of Ref. \cite{Fra85} was double-folded over the transition densities to produce form factors. Optical-model potential parameters were taken from Ref. \cite{Tos14a}.

\begin{table*}[tb]
\caption{Resonance parameters for the three-component fits. For each state with the proposed spin and parity ($J^\pi$) shown, the resonance energy ($E_r$), resonance width ($\Gamma_r$), and the population relative to the 1/2$^+$ state ($I/I_{1/2^+}$) are listed for the proton-removal reaction of Randisi {\it et al.} ($-$1p)) \cite{Ran14} as well as the present nucleon-exchange reaction  ($-$1p+1n).}
\begin{ruledtabular}
\setlength\extrarowheight{3pt}
\begin{tabular}{l|cccc|ccc}
$J^\pi$ & $E_r$ & $\Gamma_r$  & $I/I_{1/2^+}$ & \hspace*{0.3cm} & $E_r$ & $\Gamma_r$  & $I/I_{1/2^+}$ \\
 & \multicolumn{4}{c|}{Randisi {\it et al.} \cite{Ran14} ($-$1p)} & \multicolumn{3}{c}{present work ($-$1p+1n)} \\  \hline
1/2$^+$ & 0.40$\pm$0.03 & 0.80$^{+0.18}_{-0.12}$ & 1.00 & & 0.40\footnote{fixed value from Randisi {\it et al.} \cite{Ran14}} & 0.80\footnotemark[1] & 1.00 \\
5/2$^+_1$ & 0.85$^{+0.15}_{-0.11}$ & 0.30$^{+0.34}_{-0.15}$ & 0.40$\pm$0.07 & & 1.05$\pm$0.10 & 0.50$\pm$0.20 & 0.63$\pm$0.15 \\
5/2$^+_2$ & 2.35$\pm$0.14 & 1.50$\pm$0.40 & 0.80$\pm$0.09 & & 2.56$\pm$0.13\footnote{value taken from two-parameter fit} & 2.29$\pm$0.73\footnotemark[2] &  3.88$\pm$0.50 \\
\end{tabular}
\end{ruledtabular}
\label{table1}
\end{table*}

Guided by (0 -- 3)$\hbar\omega$ shell model calculations Randisi {\it et al.} analyzed their data by introducing a second lower-lying $d$-wave resonance \cite{Ran14}. The  resonance energies and widths for this analysis are listed in Table \ref{table1} together with the parameters used to fit the present data as shown in Figure \ref{Figure2}. A completely unconstrained three-resonance fit resulted in degenerate values for the lower two resonances. Thus the values for the $s$-wave resonance was constrained to the value of Randisi {\it et al.} ( $E_r$ = 0.40 MeV, $\Gamma_r$ = 0.80 MeV) and the parameters for the second $d$-wave resonance were kept at the value extracted from the two-parameter fit  ($E_r$ = 2.56 MeV, $\Gamma_r$ = 2.29 MeV). The resonance energy and width of the first $d$-wave resonance as well as strength of all three components were varied. Figure \ref{Figure2} shows that the nucleon-exchange data can be well described with parameters similar to the one-proton removal reaction.

Table \ref{table1} also includes the ratios of the $d$-wave resonances relative to the $s$-wave resonance for the two reactions. The relative intensities in the proton-removal reaction are governed by the ground state configuration of $^{14}$B where the spectroscopic factors for populating the 1/2$^+$, 5/2$^+_1$, and 5/2$^+_2$ were calculated within the WBP shell model to be 0.41, 0.13, and 0.43, respectively, in good agreement with the data \cite{Ran14}. The 1/2$^+$ and 5/2$^+_2$ states are dominated by single-particle configurations, whereas the 5/2$^+_1$ has 2$\hbar \omega$ $^{10}$Be $\otimes (\nu2s1d)^3$ parentage. 

The intensity of the low-lying $d$-wave resonance in the nucleon-exchange reaction is slightly larger than the intensity extracted from the proton-removal reaction, while the intensity of the second $d$-wave resonance is significantly larger. These ratios do not have to be the same for the two different reactions. For example, in addition to the two 5/2$^+$ states, the  (0 -- 3)$\hbar\omega$ shell model calculations also predict a low-lying 3/2$^+$ state. The spectroscopic factor of this state for proton removal from $^{14}$B is zero, so it is not expected to be observed in the data of Randisi {\it et al.} \cite{Ran14}. It could, however, be populated in the present reaction which would reduce the strengths of the two $d$-wave resonances relative to the low-lying $s$-wave resonance.  It should be mentioned that the low-lying 3/2$^+$ and 5/2$^+$ states predicted by the (0 -- 3)$\hbar \omega$ shell-model calculations using the WBP interaction \cite{Ran14} are not present in the simplified scheme by Fortune \cite{For13}. This discrepancy has recently been reiterated and is not fully understood \cite{For14}.

Finally, the present data show no evidence for any low-energy decay from the second $d_{5/2}$ to the first excited 2$^+$ state in $^{12}$Be as was suggested by Aksyutina {\it et al.} \cite{Aks13b}. Simulations including such a decay branch resulted in an upper limit of less than 10\%. This finding is consistent with results by Randisi {\it et al.} who extracted a branching ratio of 5(2)\% \cite{Ran14}.

\section{Summary and Conclusion}

In conclusion, the $^{13}$B($-$1p+1n) nucleon-exchange reaction  was used to populate the neutron-unbound nucleus $^{13}$Be. The decay-energy spectrum can be described with  resonance parameters similar to previously reported values for the proton-removal reaction from $^{14}$B. In general nucleon-exchange reactions offer an alternative reaction mechanism to selectively populate states in neutron-rich nuclei when the nucleus of interest can not be populated by single proton (i.e. $^{15}$Be, $^{20}$B, or $^{24}$N) or even two-proton ($^{23}$C) removal reactions.

We would like to thank Shumpei Noji and Remco Zegers for helpful discussions of the nucleon-exchange calculations and Paul Gueye for proof reading the manuscript. This work was supported by the National Science Foundation under Grant Nos. PHY09-69058, PHY09-69173, PHY11-02511, PHY12-05537, and PHY13-06074. This material is based upon work supported by the Department of Energy National Nuclear Security Administration under Award Number DE-NA0000979.



\end{document}